\documentclass[rmp,twocolumn]{revtex4-1}
\usepackage{graphicx,amsmath,amssymb}
\usepackage{color}
\usepackage[normalem]{ulem}	

\begin{document}

\title{Physics picture from neutron scattering study on Fe-based superconductors}
\author{Wei Bao}
\affiliation{Department of Physics, Renmin University of China, Beijing
100872, China }

\begin{abstract}
Neutron scattering, with its ability to measure the crystal structure, the magnetic order, and the structural and magnetic excitations, plays an active role in investigating various families of Fe-based high-T$_c$ superconductors. Three different types of antiferromagnetic orders have been discovered in the Fe plane, but two of them cannot be explained by the spin-density-wave (SDW) mechanism of nesting Fermi surfaces. Noticing the close relation between antiferromagnetic order and lattice distortion in orbital ordering from previous studies on manganites and other oxides, we have advocated orbital ordering as the underlying common mechanism for the structural and antiferromagnetic transitions in the 1111, 122 and 11 parent compounds.
We observe the coexistence of antiferromagnetic order and superconductivity in the (Ba,K)Fe$_2$As$_2$ system, when its phase separation is generally accepted. Optimal T$_c$ is proposed to be controlled by the local FeAs$_4$ tetrahedron from our investigation on the 1111 materials. The Bloch phase coherence of the Fermi liquid is found crucial to the occurrence of bulk superconductivity in iron chalcogenides of both the 11 and the 245 families. Iron chalcogenides carry a larger staggered magnetic moment ($>2\mu_B$/Fe) than that in iron pnictides ($<1\mu_B$/Fe) in the antiferromagnetic order. Normal state magnetic excitations in the 11 superconductor are of the itinerant nature while in the 245 superconductor the spin-waves of localized moments. The observation of superconducting resonance peak provides a crucial piece of information in current deliberation of the pairing symmetry in Fe-based superconductors.

\end{abstract}


\maketitle

\tableofcontents

\section{Introduction}

In 2008, the research group led by Hosono reported the discovery of superconductivity at 26 K in LaFeAs(O,F) \cite{Kamihara2008} after having discovered superconductivity below 4 K in isostructural LaFeP(O,F) \cite{Kamihara2006} and LaNiP(O,F) \cite{Watanabe2007}. Superconductivity at comparable T$_c$ has been reported in YNiBC \cite{YNiBC} and related materials \cite{YPdBC} of a laminar crystal structure similar to LaFeAsO \cite{NiBC_str} in 1990s. Upon replacing the nonmagnetic Y by magnetic rare-earth elements, T$_c$ of these borocarbide superconductors is suppressed in accordance to the expectation for the phonon-mediated $s$-wave superconductors \cite{bswave}.
When La is replaced by magnetic Sm or Ce in the LaFeAs(O,F) superconductor, however,
T$_c$ increases to above 40 K \cite{A033603,A033790}. The highest T$_c \approx 55$ K of the Fe-based superconductors is also achieved in the Sm compound \cite{A042053}. The absence of the $s$-wave pair-breaking by local magnetic ions, therefore, strongly suggests an unconventional superconductivity in the Fe-based materials. The rush to the ``iron age'' of high-T$_c$ superconductors is henceforth unleashed \cite{rev2009H}.

Thousands of papers have been published on Fe-based superconducting materials and comprehensive reviews already exist \cite{rev2009H,rev2010l,rev2011l}. Here the scope is limited to physics picture derived mainly from our neutron scattering investigation.
Personal viewpoints are expressed in order to stimulate discussion in the active and evolving field. We also limit ourselves to a sketchy outline to emphasize the main points. Original publications should be consulted for details. 

\section{Orbital ordering: the common mechanism for the structural and antiferromagnetic transitions}

Band-structure theory provides the first glimpse of electronic structure in the Fe-based
superconductors \cite{bs07,A030429,A033325,A033236,A033286}.
A preeminent feature is the pair of quasi-two-dimensional hole and electron Fermi surfaces that are mostly composed of $d$-orbitals of Fe. This is supported by ensuing ARPES measurements from various groups. The nesting of the hole and electron Fermi surfaces would naturally lead to a spin-density-wave (SDW) antiferromagnetic order \cite{A033236,A033286,A033426,A034346}. Neutron diffraction observation of an antiferromagnetic order at the predicted in-plane wave-vector \cite{A040795} has been hailed as the proof of the SDW theory.

However, the same neutron diffraction work uncovers that the expected SDW gap in transport property \cite{A034346} occurs not at the antiferromagnetic transition, but at a separated structural transition at a higher temperature T$_S$ \cite{A040795}. This important fact cannot be explained by the SDW theory and, unfortunately, has been ignored by most studies in the field. The SDW mechanism is a well-established way for a weak-coupling metal to acquire antiferromagnetic long-range order, such as in Cr \cite{sdw_lom,Cr_rev}. However, experimental results of the Fe-based materials accumulated over time have indicated the necessity of supplementing it by localized spin features such as orbital ordering. 

A monoclinic $P112/n$ unit cell, which is easier to compare with the ZrCuSiAs (1111) $P4/nmm$ tetragonal unit cell, was used for the distorted lattice below T$_S$ in the initial neutron diffraction work on LaFeAsO \cite{A040795}. Since then, the consensus space group to describe the low-temperature structure has been the orthorhombic $Cmma$ \cite{A043569}.
The antiferromagnetic structure in the orthorhombic unit cell is determined for the first time in the neutron diffraction work on NdFeAsO \cite{A062195}. The antiferromagnetic wave-vector is (100) below 2 K and the easy-axis of the Fe moment is also along the elongated $a$-axis in the basal plane. The staggered moment is 0.9(1)$\mu_B$ per Fe at 0.3 K. 

When the Nd moments enter the paramagnetic state above 2 K, the 
antiferromagnetic wave-vector of the Fe sublattice is switched to (10$\frac{1}{2}$) with a much reduced staggered moment of 0.25(7)$\mu_B$ per Fe \cite{A070662}, which has been overlooked in our previous work \cite{A062195}. The Nd antiferromagnetic order changes the Fe inter-layer magnetic coupling from antiferromagnetic to ferromagnetic and enhances the Fe staggered moment. However, it does not affect the in-plane Fe moment arrangement. The same in-plane Fe antiferromagnetic structure as that in NdFeAsO with both the antiferromagnetic wave-vector and moment easy-axis along the $a$-axis in basal plane \cite{A062195} is later confirmed also for CeFeAsO \cite{A062528}, PrFeAsO \cite{A074441,A074872} and LaFeAsO \cite{A094816} with neutron diffraction. Note that the perpendicular depiction of the Fe moment and the antiferromagnetic wave-vector in Fig.~4 of Ref.\ \cite{A040795} needs to be rectified.

In BaFe$_2$As$_2$, the resistivity anomaly does concur with antiferromagnetic transition as expected by the SDW theory, however, so does the structural transition deforming from the tetragonal $I4/mmm$ ThCr$_2$Si$_2$ (122) structure \cite{A062776}. As in the 1111 parent compounds, both the in-plane component of the antiferromagnetic wave-vector (101) and the Fe easy-axis are along the elongated $a$-axis of the orthorhombic $Fmmm$ unit cell. The staggered moment is 0.87(3)$\mu_B$ per Fe at 5 K. Almost identical antiferromagnetic order is also found in SrFe$_2$As$_2$ \cite{A071077} and CaFe$_2$As$_2$ \cite{A071525}.

The Fe moments in both the 1111 and the 122 families of parent compounds align parallel to each other between the shorten nearest-neighbor (n.n.) pairs along the $b$-axis, and they align antiparallel between the elongated n.n.\ pairs along the $a$-axis. Namely,
the antiferromagnetic bond shrinks and the ferromagnetic bond expands in the FeAs plane in both families. One might argue that the SDW mechanism drive the 122 system to an antiferromagnetic transition and the magnetostriction leads to the lattice distortion at the same time. However, as mentioned above, the lattice distortion predates the antiferromagnetic transition in the 1111 system. Thus it cannot be caused by magnetostriction during a magnetic transition. 

Alternatively, if one assumes an orbital ordering that realizes preferred occupation of one $d$-orbital in the $a$-axis and another $d$-orbital in the $b$-axis, it could naturally lead to expanded antiferromagnetic bonds in one in-plane direction and shortened ferromagnetic bonds in the other. We have advocated the $d_{xz}$ and $d_{yz}$ orbitals for such differentiated roles since our NdFeAsO work \cite{A062195} together with the {\it ab initio} work by one of our co-workers \cite{A042252}.
Historically, Goodenough \cite{orb_ge} come up with different orbital orders to form  orbitally ordered lattices, which leads to the desired super-exchange or double-exchange interactions between spin pairs, to account for various magnetic orders observed in neutron diffraction work on manganites by Wollan and Koehler \cite{wollan}. 
For review on recent studies, refer to \cite{rev_mit}.

When orbital ordering occurs, a structural transition will appear, which will generally establish a new spin Hamiltonian \cite{bao96b,bao96c}. When the magnetic transition temperature of the new spin Hamiltonian is lower than the orbital transition temperature, magnetic transition will occur after the structural transition upon cooling. We have investigated such a behavior in our previous neutron scattering study on manganites \cite{bao96b}, and
the 1111 family of parent compounds belong to this type. When the magnetic transition temperature of the new spin Hamiltonian is higher than the orbital transition temperature, magnetic transition will occur in a first-order transition as soon as the structural transition occurs. We have also encountered such a behavior in our previous neutron scattering study on vanadium sesquioxides \cite{bao96c}, and
the 122 family of parent compounds belong to this type \cite{A062776}. 

Also belonged to the latter type are the 11 family of parent compounds Fe$_{1+y}$(Te$_{1-x}$Se$_x$) \cite{A092058}, in which magnetic and structural transitions occur simultaneously. Here, the SDW theory completely fails to explain the observed commensurate and incommensurate antiferromagnetic order, with the predicted antiferromagnetic wave-vector 45$^o$ away from the observed ($\delta$0$\frac{1}{2}$). The commensurate antiferromagnetic structure of $\delta=\frac{1}{2}$ occurs for $y< 0.08$ with a monoclinic $P2_1/m$ unit cell in metallic state, while the incommensurate one of $\delta<\frac{1}{2}$ occurs for $y> 0.08$ with an orthorhombic $Pmmn$ unit cell in semiconducting state. The amplitude of the staggered moment in both the commensurate and incommensurate structures is 2$\mu_B$ per Fe, with an appreciable canting component in a generally rather complex magnetic structure. 
Nevertheless, the same relation between (anti-) ferromagnetic spin pair and (expanded) shortened bond length is observed, further supporting a unified orbital ordering scenario for all the three families of Fe-based superconductors \cite{A092058}.

The lattice Fourier transformation of the magnetic exchange constants with the signs described above which are prepared by the orbital ordering $J({\bf Q})$, therefore, will peak at the observed antiferromagnetic wave-vector \cite{A062195,A062776,A092058}, while the bare Lindhard function $\chi_0({\bf Q},\omega=0)$ always peaks at the same nesting wave-vector in the 1111, 122 and 11 parent compounds \cite{A033236,A033286,A033426,A034346}. In the RPA approximation, the magnetic susceptibility is
\begin{equation}
\chi({\bf Q},\omega)=\frac{\chi_0({\bf Q},\omega)}{1-J({\bf Q})\chi_0({\bf Q},\omega)}.
\end{equation} 
Thus, one can image that in the 1111 and the 122 families, the orbital ordering and the Fermi surface work in step in realizing the observed antiferromagnetic structure, since both the $J({\bf Q})$ and $\chi_0({\bf Q},\omega=0)$ peak around the nesting wave-vector. In the 11 family, however, they work out of step. The $J({\bf Q})$ peaking at $(\delta,0)$ dominates over the $\chi_0({\bf Q},\omega=0)$ and moves the peak of $\chi({\bf Q},\omega=0)$ from the nesting wave-vector to the observed $(\delta,0)$.

The commensurate antiferromagnetic order in the 11 parent compounds has been investigated previously in 1975 by Fruchart et al. \cite{FeTe_comm} and also in a later work \cite{A110195}. The staggered moment in \cite{A110195} is similar to that found in our work \cite{A092058}. Note that the appreciable canting component observed in experiments is often neglected in popular ``bi-linear'' depiction of the commensurate antiferromagnetic order.

The electronic hopping, either real in determining the double-exchange or virtual in the 
super-exchange, should be reflected in transport \cite{orb_ge,rev_mit}. Since magnetic exchange constants in the $a$-axis and $b$-axis have different signs, in-plane anisotropy in transport properties is expected
and data have begun to appear since 2010 in measurements of detwinned 122 parent compounds \cite{C023801}. The static measurement of the in-plane transport anisotropy has been extended later to frequency up to 2eV in optical spectroscopic study on detwinned BaFe$_2$As$_2$ \cite{D064967}. The preferred occupation of the $d_{xz}$ orbital at $k_z\sim \pi$ observed in polarization-dependent ARPES study on BaFe$_2$As$_2$ offers a more direct evidence for orbital ordering \cite{B041632}.
 
In addition to the theoretical study of Yildirim \cite{A042252}, there has been persistent albeit small theoretical pursuit on orbital ordering in the Fe-based superconductors \cite{B051704,B052957,B101573,C044611,E081917}. The recently observed block antiferromagnetic order in vacancy-ordered $A_2$Fe$_4$Se$_5$ \cite{D020830,D022882}, to be presented later, can also be understood together with those observed in the 1111 \cite{A062195}, 122 \cite{A062776} and 11 \cite{A092058} compounds in a unified orbital ordering scenario \cite{D050432,E060881}.

\section{Structural control of superconductivity}

Superconductivity is achieved by doping the 1111 and 122 parent compounds to suppress the structural and antiferromagnetic transitions \cite{Kamihara2008,A054630}. Doping with charges of opposite sign \cite{Wen2008,A072223,A072237}, isovalent substitution \cite{B013227} and high pressure or uniaxial stress \cite{A070616,A112554,B050968} also work.
The FeSe \cite{A072369} and LiFeAs \cite{A064688} do not require intentional doping, but the samples are nonstoichiometric and already in the $P4/nmm$ tetragonal PbO (11) or PbFCl (111) structure. Therefore, antiferromagnetic order seems to be a competing phase to superconductivity. Doping, chemical pressure, and physical pressure aim to suppress the structure distortion and its accompanying antiferromagnetic order.

Superconducting phase is dome-like in the tetragonal phase for the 1111 and 122 Fe-based superconductors. There have been several proposals concerning which material parameter is crucial in controlling the superconductivity. Lee et al.~\cite{A063821} and Qiu et al.~\cite{A062195} have proposed in 2008 that {\em the closer the As-Fe-As angle is to the ideal tetrahedral angle} $\arccos\left(-\frac{1}{3}\right)= 109.47^o$, {\em the higher the T$_c$ is}.

This local structural mechanism has survived since the initial studies on the 1111 system.
It obviously links closely with the orbital ordering scenario for the antiferromagnetic order. A less perfect Fe-As tetrahedron environment would lift the $d$ orbital degeneracy and differentiate the $d$ orbitals, thus favor the antiferromagnetic instability.

\section{Coexistence of superconductivity and antiferromagnetism}

While antiferromagnetic order and superconductivity look like competing order parameters, an important fine point is whether they can coexist in part of the phase diagram.

For the 1111 system, competing claims exist. On the one hand, antiferromagnetic phase and the superconducting phase are mutually exclusive, such as in the investigations on the Ce  \cite{A062528} and La systems \cite{A063533,A094816}. One the other hand, antiferromagnetic phase and the superconducting phase can coexist, such as in the studies on the La \cite{A064798} and Sm \cite{A074876} system.
It is not clear whether the difference is real or is caused by the difficulty in sample synthesis control, given the narrow composition region where the difference in claims occurs.

To address the issue, we investigate the phase diagram of the K-doped BaFe$_2$As$_2$ system. The system has the merit that the whole composition range from Ba to K can be synthesized \cite{A061301}. By combining neutron diffraction, synchrotron x-ray diffraction and bulk measurements, we find that the simultaneous magnetic and structural transition of BaFe$_2$As$_2$ is suppressed by K doping. However, the superconducting phase emerges before the antiferromagnetic order is completely suppressed. We establish that the coexistence of antiferromagnetic phase and the superconducting phase is an intrinsic property of the system \cite{A073950}.

There are other views. The first is that the structural and antiferromagnetic transitions are separated ones. We found in the case of BaFe$_2$As$_2$ that the separation is likely caused by flux-contaminated single-crystal samples used \cite{B010738}. Our conclusion of the simultaneous magnetostructural transition has been confirmed by a recent diffraction work using high quality (Ba,K)Fe$_2$As$_2$ samples \cite{D021933}.

The second claim is that the samples phase-separate into an antiferromagnetic phase and a superconducting phase, instead of coexisting as shown in our study. It is understandable that one fails to make homogeneous samples when a new material first appears before the correct recipe is widely known. However, these two claims due to poor sample quality have been so widely propagated that they have permeated many otherwise respectable reviews and perspective pieces, e.g. in \cite{rev2010n}. The coexistence of antiferromagnetism and superconductivity in (Ba,K)Fe$_2$As$_2$ has been corroborated using a different experimental technique in recent NMR studies \cite{C053718,E042434}. 

We used nominal sample composition in our study \cite{A073950}. Rotter et al.\ did a better job in determining the sample composition in a slightly later study \cite{A074096}. However, they did not investigate the antiferromagnetic phase transition and the coexistence problem. Avci et al.\ refine the sample compositions in their systematic work \cite{D021933}.

Since our study on the hole-doped phase-diagram using the (Ba,K)Fe$_2$As$_2$ system as the prototype, phase-diagrams of electron-doping using Co or other transition metal elements at the Fe site, and isovalent substitution of As by P \cite{B013227} have also been established to show the coexistence as a general feature in the 122 system. Pressure phase-diagram has also been obtained but it is extremely sensitive to the hydrostatic condition \cite{A112554,B050968}, not completely surprising for their two-dimensional crystalline structures.

Theoretical consequences of the coexistence have been reviewed recently \cite{rev2011t}.

\section{Weak localization and superconductivity}

The discovery of the 11 superconductors expands the scope of Fe-based high-T$_c$ superconductors from pnictides to less poisonous chalcogenides \cite{A072369}.
The original 11 superconductor contains more iron than selenium, and it was initially considered to form in the PdO structure with Se deficiency. We refine the isostructural iron telluride and find that it forms in the PbFCl (111) structure of the same space group with the excess Fe(2) occupying the same site as the Li in LiFeAs \cite{A092058}.
This is also the case found later for iron selenide \cite{A111613}.
Thus the customary classification of 11 and 111 families of Fe-based superconductors is less structural than electronic in balancing the Se$^{-2}$ and As$^{-3}$ ions respectively.

The isovalent substitution of Se by Te has not affected T$_c\approx 8$ K much until a first-order jump to the T$_c\approx 14$ K phase \cite{A074775,A080474}. Close to the Te end of the phase diagram, there exists an antiferromagnetic phase in a narrow composition range\cite{C035647}. The magnetic structure, crystal symmetry, and transport property are different in the two phases at the Te end tuned by the amount of the excess Fe \cite{A092058}, which has been presented above. What we concern here is the strong spin-glass low-energy magnetic fluctuations left in the samples after the long-range antiferromagnetic order is suppressed by Se substitution of Te.

It turns out that four n.n.\ spins on the Fe square lattice in the tetragonal $P4/nmm$ phase tend to form a super-spin block, and these block spins then form a chessboard antiferromagnetic pattern. However, this tendency somehow cannot develop into a long-range order, and a short-range order is instead nucleated by the excess Fe and becomes spin-glass at low temperature \cite{D095196}. On the other hand, magnetic entropy is released near the Te end through the usual route of structural transition for frustrated magnetic systems: The new spin Hamiltonian prepared by the orbital ordering allows magnetic transition at higher temperature.

We have shown that the more the excess Fe is, the stronger the low-energy short-range antiferromagnetic fluctuations are. When the magnetic fluctuations are strong enough, their scattering turns metallic resistivity to logarithmic divergence \cite{C035647}. Bulk superconductivity in
Fe(Te,Se) occurs only when normal state transport is metallic. Therefore, when low-energy scattering leads to weak localization to break the Fermi liquid phase coherence of the Bloch wave function, superconductivity is destroyed. 

As will be presented, a variation of this behavior happens in the Fe-vacancy ordered iron chalcogenide superconductors.

\section{Structural and magnetic excitations}

Electron-phonon interaction plays a central role in the BCS theory of superconductivity.
When triple-axis inelastic neutron scattering spectrometer was invented, its major application was to measure the spectrum of phonon. Theoretical and experimental studies together have propelled {\it ab initio} calculation of phonon spectrum to a very high reliability for most materials. When the discovery of the 1111 Fe-based high T$_c$ superconductivity was reported, theoretical phonon spectrum was used to calculate T$_c$
 and it was concluded that phonon are not energetic enough to support the high value of T$_c$ \cite{A032703}.

To check the reliability of the conclusion, we try to verify the energy scale of the theoretic phonon spectrum in the La-1111 superconductor (T$_c =26$~K) \cite{A051062}. This can be readily performed by comparing the peak energy due to van Hove singularity in phonon density of states with the theoretic value, and the theory is indeed found to be reliable. The check was repeated with coarser energy resolution to cover the whole phonon band and the same conclusion is reached \cite{A073370}. Studies on phonon density of states for 
BaFe$_2$As$_2$ \cite{A073172} and FeSe \cite{Phelan} have also been performed.

Meanwhile, theories for the magnetic origin of the high-T$_c$ superconductivity in iron pnictides are advanced. Similar to the case for cuprate superconductors, $d$-wave was a popular choice. Thus, it was a big surprise that an isotropic superconducting gap was announced in an Andreev reflection study on SmFeAsO$_{0.85}$F$_{0.15}$ \cite{A054616}.
The nodeless superconducting gaps are demonstrated shortly by ARPES measurement of Ba$_{0.6}$K$_{0.4}$Fe$_2$As$_2$ \cite{A070419}. The focus then shifts to the $s^{\pm}$ symmetry for the Cooper pairs \cite{A032740}.

ARPES cannot detect the sign change from the electron to hole Fermi surface. However, the sign change will lead to a ``resonance peak'' which can be measured by inelastic neutron scattering \cite{A041793,B030008}. We try but fail to detect such a resonance peak in an optimally doped LaFeAsO$_{0.87}$F$_{0.13}$ polycrystalline sample \cite{A051062}, possibly because the superconducting volume fraction is too small. The first successful observation of the resonance peak is from a polycrystalline sample of Ba$_{0.6}$K$_{0.4}$Fe$_2$As$_2$ \cite{A073932}. When large single crystal samples of the 122 system become available,
the spatial distribution of the resonance peak can be investigated and its two-dimensionality is convincingly demonstrated \cite{A114755}.

After the discovery of the 11 superconductors, large single crystals of the 14 K Fe(Te,Se) superconductors are soon available. In addition to the expected resonance peak, it is shown that the magnetic excitations are fully gapped in the superconducting state \cite{B053559}. While the intensity of the resonance peak decreases with the rising temperature in the BCS fashion, the peak energy itself is observed to be temperature independent. The spectral weight of the resonance peak comes from gapping the inclined normal state spin excitation continuum originating from a pair of perpendicular incommensurate points \cite{B114713}. Here, the normal state magnetic excitation spectrum is entirely of a typical itinerant antiferromagnet to which spin-wave theory is not applicable \cite{bao96a}. No ``hour-glass'' type of excitations are present in our data and we suspect that it is due to superposition 
of signals from superconducting and non-superconducting portions contained in inhomogeneous samples.

There have been reports of resonance peak observed in other polycrystalline 1111 superconductors. However, the signal to noise ratio seems too low to be conclusive at this time. More recently, quality single crystals of the NaFeAs superconductor are successfully grown. Dai et al.\ have observed beautiful resonance peak from the samples to be published. 

The search for superconducting resonance peak in Fe-based superconductors has been inspired
by the theory for the $s^{\pm}$ symmetry \cite{A032740,A041793,B030008} and it has fulfilled the theoretical expectation. However, it has been recently pointed out that the $s^{\pm}$ pairing function is not a parity eigenstate when the full local symmetry of the FeAs or FeSe layer is considered. The neutron resonance data would support an odd parity state over the even parity state in the new theoretical scheme \cite{F032624}.

Different from neutron scattering investigation on magnetic excitations in Fe-based superconductor, such investigation in cuprate superconductors proceeded from parent compounds to superconducting compounds. First of all, quality samples suitable for inelastic neutron scattering study became available initially only for the parent cuprates.
Secondly, magnetic excitations from parent insulating antiferromagnet are spin-waves and they are theoretically easy to be analyzed. For iron pnictides, even the parent compounds are metallic and itinerant antiferromagnetic excitations are important. Even so, at least the lower energy parts of the magnetic excitations have been reported to be spin-waves for some parent compounds and they are readily analyzable \cite{B013784}. A recent review of experimental and theoretical results can be found in Ref.\ \cite{rev2012p}. However, as mentioned above, for the 11 superconductors, spin-wave theory completely fails \cite{B114713}.

\section{The 245 superconductors}

In 2010, a 30 K superconductor of the nominal composition K$_{0.8}$Fe$_2$Se$_2$ is reported \cite{C122924}. The chemical formulas indicates a deep electronic doping and ARPES data are consistent with such an electron counting \cite{C125980}.
However, the single crystal samples grown with the same method by the same group as those used in the ARPES experiment are shown by x-ray single crystal diffraction to be K$_{0.775(4)}$Fe$_{1.613(1)}$Se$_2$ and Cs$_{0.748(2)}$Fe$_{1.626(1)}$Se$_2$ \cite{D014882}.
Another superconducting single crystal supplied by the second group is refined to be 
K$_{0.737(6)}$Fe$_{1.631(3)}$Se$_2$. The superconducting sample of slightly different nominal composition from the second group was used in neutron powder diffraction study, and the sample composition is refined to be K$_{0.83(2)}$Fe$_{1.64(1)}$Se$_2$ \cite{D020830}.

While we cannot determine whether K or Cr is ordered in the crystal, Fe vacancies form an almost perfect $\sqrt{5}\times\sqrt{5}$ superlattice in the superconducting samples \cite{D020830,D014882}.
The refined chemical formulas is close to $A_{0.8}$Fe$_{1.6}$Se$_2$. It can be rewritten as $A_{2}$Fe$_{4}$Se$_5$. Thus, we call the new family of superconductors as the 245 superconductors.

Meanwhile, Bacsa et al.\ report in an x-ray single crystal diffraction study that two {\em non}-superconducting samples are refined to be K$_{0.93(1)}$Fe$_{1.52(1)}$Se$_2$ and 
K$_{0.862(3)}$Fe$_{1.563(4)}$Se$_2$ \cite{D020488}. At the measurement temperature of 100 and 90 K respectively, the two samples also show the $\sqrt{5}\times\sqrt{5}$ superlattice order. However, the departure from the ideal K$_{0.8}$Fe$_{1.6}$Se$_2$ formulas necessitates the partial filling of the ``vacant'' Fe2 site and a deficiency in the ``full'' Fe1 site.

Such an imperfect $\sqrt{5}\times\sqrt{5}$ Fe vacancy order exists in an intermediate temperature range even in 
non-superconducting sample K$_{0.99}$Fe$_{1.48}$Se$_2$ \cite{D023674}. Close to be KFe$_{1.5}$Se$_2$, namely one vacancy out of four Fe sites, the perfect $2\sqrt{2}\times\sqrt{2}$ pattern occurs only in part of the sample in the temperature window of from 295 to 500 K. This can be understood in the two-level statistical physics model provided that the 
$\sqrt{5}\times\sqrt{5}$ lattice structure is lower than the $2\sqrt{2}\times\sqrt{2}$ structure in energy. In calculating these energies, the observed block antiferromagnetic order and its associated lattice tetramerization \cite{D020830} should be included in the equation \cite{D021344,D022215}.

All these samples can be written as K$_{x}$Fe$_{2-y}$Se$_2$ with $x\sim 2y$. Namely, with K$^{+1}$ and Se$^{-2}$, valence of Fe $\sim 2$ \cite{D014882,D020488,D020830,D023674}. The phase-diagram is provided in \cite{D023674}. Note that the order-disorder structural transitions occur at T$_S$ and T$^*$, and it is remarkable that Fe ions remain mobile at such  low temperatures. It is useful to keep these information in mind if one want to image the element movement during the sample annealing process.

Among the discoverers of the 245 superconductors, Fang et al.\ distinguishes themselves in intentionally introducing Fe vacancy into the Fe square lattice, as reflected in the original arXiv article title \cite{C125236,D010462}. The T$^*$ approaches zero in the (Tl,K)Fe$_x$Se$_2$ system \cite{C125236} which resembles the usual magnetic susceptibility behavior at T$_N$ \cite{D023674}, but the anomaly in magnetic susceptibility is caused by the large magnetic excitation gap due to the moment-space anisotropy in the $\sqrt{5}\times\sqrt{5}$ phase \cite{E012413}. For all the five members of the 245 superconductors, T$_N$ stays high above room temperature, ranging from 471 to 559 K \cite{D022882}. The strong interaction between the antiferromagnetic and superconducting orders is reflected in the anomaly in magnetic order parameter near T$_c$ \cite{D020830,D022882}.

The imperfect $\sqrt{5}\times\sqrt{5}$ vacancy order will contribute to impurity scattering to conducting electrons. Resistivity traces the degree of the order well, changing from semiconducting to logarithmic divergence and then to metallic when the sample composition approaches the ideal 245 formulas \cite{D023674}. Like in the 11 superconductors, superconductivity occurs when the normal state transport is metallic with minimum scattering from imperfect vacancy order.

Different from all previous antiferromagnetic orders found in the Fe-based superconductor systems, antiferromagnetic order in the 245 superconductors does not break the four-fold symmetry of the $I4/m$ lattice structure \cite{D020830}. Differentiation of magnetic exchange interactions therefore is not in the $a$-axis or $b$-axis direction but in the intra or inter four-spin blocks. Magnetic excitations of the (Tl,Rb)$_2$Fe$_4$Se$_5$ superconductor have been measured recently and the spin-wave theory including the n.n.\ and n.n.n.\ intra and inter-block exchange constants well describes the data \cite{E012413}. The success of the spin-wave theory is not surprising given that the staggered moment 3.3$\mu_B$ per Fe is very close to the atomic limit value 4$\mu_B$.

With the volatile alkali element and highly mobile Fe, the $A_{x}$Fe$_{2-y}$Se$_2$ crystal often contains plate-like intergrowth of Fe \cite{D020830}. Thus, extra Fe is included in the nominal sample composition determined by inductively coupled plasma analysis. Another common error is  identifying the 245 phase with the spotting of the $\sqrt{5}\times\sqrt{5}$ superlattice pattern using surface sensitive method such as TEM or STM. Even detecting the $\sqrt{5}\times\sqrt{5}$ pattern with bulk probe such as neutron scattering cannot identify the sample as $A_2$Fe$_4$Se$_5$ and we have provided the non-superconducting refinement examples K$_{0.99}$Fe$_{1.48}$Se$_2$, K$_{0.93(1)}$Fe$_{1.52(1)}$Se$_2$ and 
K$_{0.862(3)}$Fe$_{1.563(4)}$Se$_2$ above \cite{D020488,D023674}.
Many samples in the $I4/m$ insulator phase of the phase-diagram shown in Fig.\ 4 of Ref. \cite{D023674} have been misidentified as $A_2$Fe$_4$Se$_5$ in the current literature.

\section*{Acknowledgment}

We thank all of our collaborators in the investigations on Fe-based superconductors, in particular Drs. X.H. Chen, M.H. Fang, Z.Q. Mao, Y.M. Qiu, Q.Z. Huang, T. Yildirim, C. Broholm, A.T. Savici, D.N. Argyriou, and F. Ye.  
This work was supported by the National Basic Research Program of China (Grant Nos.\ 2012CB921700, and 2011CBA00112) and  
the National Natural Science Foundation of China (Grant Nos.\ 11034012, and 11190024).


%

\end{document}